\documentclass[runningheads]{llncs}
\usepackage[T1]{fontenc}

\usepackage{graphicx}
\usepackage{amsmath}
\usepackage{amssymb}
\usepackage{amsfonts}
\usepackage{booktabs}
\usepackage{pifont}
\usepackage{multirow}
\usepackage{enumitem}
\usepackage{flushend}
\usepackage{tabularx}
\usepackage{orcidlink}

\usepackage[capitalize]{cleveref}
\crefname{section}{Sec.}{Secs.}
\Crefname{section}{Section}{Sections}
\Crefname{table}{Table}{Tables}
\crefname{table}{Tab.}{Tabs.}

\usepackage{glossaries}
\newacronym{asvspoof}{ASVspoof}{Automatic Speaker Verification And Spoofing Countermeasures}
\newacronym{ser}{SER}{Speech Emotion Recognition}
\newacronym{pa}{PA}{Physical Access}
\newacronym{la}{LA}{Logical Access}
\newacronym{df}{DF}{deepfake}
\newacronym{asv}{ASV}{Automatic Speaker Verification}
\newacronym{tts}{TTS}{Text-to-Speech}
\newacronym{vc}{VC}{Voice Conversion}
\newacronym{tdnn}{TDNN}{Time Delay Neural Network}
\newacronym{svm}{SVM}{Support Vector Machine}
\newacronym{eer}{EER}{Equal Error Rate}
\newacronym{roc}{ROC}{Receiver Operating Characteristic}
\newacronym{auc}{AUC}{Area Under the Curve}
\newacronym{gru}{GRU}{Gated Recurrent Unit}
\newacronym{ai}{AI}{Artificial Intelligence}
\newacronym{mfcc}{MFCC}{Mel-frequency Cepstral Coefficient}
\newacronym{dl}{DL}{Deep Learning}
\newacronym{stft}{STFT}{Short Time Fourier Transform}
\newacronym{gan}{GAN}{Generative Adversarial Network}
\newacronym{cnn}{CNN}{Convolutional Neural Network}
\newacronym{pca}{PCA}{Percentage of Correct Attribution}
\newacronym{ljs}{LJS}{LJSpeech}
\newacronym{po}{PO}{Amazon AWS Polly}
\newacronym{gs}{GS}{Google Cloud Standard}
\newacronym{gw}{GW}{Google Cloud WaveNet}
\newacronym{az}{AZ}{Microsoft Azure}
\newacronym{wa}{WA}{IBM Watson}
\newacronym{au}{AU}{authentic}
\newacronym{iemocap}{IEMOCAP}{Interactive Emotional Dyadic Motion Capture}

\usepackage{color}

\begin{document}

\title{Combining Automatic Speaker Verification and Prosody Analysis for Synthetic Speech Detection}

\titlerunning{Combining ASV and Prosody Analysis for Synthetic Speech Detection}

\author{Luigi Attorresi\inst{1}\orcidlink{0000-0002-0180-1323} \and
Davide Salvi\inst{1}\orcidlink{0000-0002-5163-3364} \and
Clara Borrelli\inst{1}\orcidlink{0000-0002-8127-9976} \and
Paolo Bestagini\inst{1}\orcidlink{0000-0003-0406-0222} \and
Stefano Tubaro\inst{1}\orcidlink{0000-0002-1990-9869}}

\authorrunning{Attorresi et al.}

\institute{Dipartimento di Elettronica, Informazione e Bioingegneria\\ 
Politecnico di Milano, Milan, Italy\\
\email{luigi.attorresi@mail.polimi.it, \{davide.salvi, clara.borrelli, paolo.bestagini, stefano.tubaro\}@polimi.it}}

\maketitle              

\begin{abstract}
The rapid spread of media content synthesis technology and the potentially damaging impact of audio and video deepfakes on people's lives have raised the need to implement systems able to detect these forgeries automatically.
In this work we present a novel approach for synthetic speech detection, exploiting the combination of two high-level semantic properties of the human voice.
On one side, we focus on speaker identity cues and represent them as speaker embeddings extracted using a state-of-the-art method for the automatic speaker verification task.
On the other side, voice prosody, intended as variations in rhythm, pitch or accent in speech, is extracted through a specialized encoder.
We show that the combination of these two embeddings fed to a supervised binary classifier allows the detection of deepfake speech generated with both Text-to-Speech and Voice Conversion techniques.
Our results show improvements over the considered baselines, good generalization properties over multiple datasets and robustness to audio compression.

\keywords{Synthetic Speech Detection \and Deepfake \and Audio Forensics \and Prosody \and Speaker Verification}
\end{abstract}

\glsresetall
\section{Introduction}
\label{sec:introduction}

The term \gls{df} refers to a category of synthetic multimedia content generated through \gls{dl} techniques that depict individuals in actions and behaviors that do not belong to them.
In recent years, the fast development in this technology has made it increasingly realistic and accessible. This enables producing manipulated media that are almost impossible to distinguish from the original ones~\cite{fakenewscientist}.
These improvements result in exciting and futuristic scenarios but also represent a potential tool for malicious purposes~\cite{de2021distinct,westerlund2019emergence}.
There are several cases where \gls{df} videos have been employed in the creation of non-consensual adult material~\cite{deepfakesforbes}, fake political news~\cite{deepfakeguardian} or damage people's reputation~\cite{deepfakenyt}. 
Likewise, with the increasing quality and accessibility of speech synthesis techniques, namely \gls{tts} and \gls{vc}, \gls{df} voices have emerged and proved equally dangerous. The difference between these techniques is the starting point of the synthesis process, which is text for \gls{tts} and voice for \gls{vc}.
Both of them proved to be capable of fooling recognition systems into accessing the victim's personal information and committing frauds~\cite{fraudsterforbes} or by providing support for voice phishing attacks~\cite{deepfakemimecast}.

Given the threat posed by these technologies, there is an urgent need to develop systems able to detect their misbehaving use.
Several state-of-the-art methods have been proposed to face this problem for both videos and audio recordings~\cite{verdoliva2020media,masood2021deepfakes,kamble2020advances,bonettini2020video,chen2020generalization}.
These can be divided into two main groups.
The first one includes methods that focus on low-level characteristics of the signal~\cite{zhang2019detecting,li2018exposing,gao2021generalized}, looking for artifacts introduced by the generators at the pixel or sample level. These artifacts are interpreted as hidden fingerprints left by the synthesis process that we can leverage to determine the authenticity of a given media content.
For example, the method proposed in~\cite{malik2019securing} detects synthesized speech by looking for artifacts through high-order spectral analysis. It performs quadrature-phase coupling in the estimated bicoherence and a series of test statistics for Gaussianity and linearity, assuming that an authentic recording has higher non-linearity than a counterfeited one.
Similarly, \cite{borrelli2021synthetic} addresses the same problem by combining a set of features that model speech as an auto-regressive process and evaluating the effect of including bicoherence as well, which proved useful in \cite{agarwal2021detecting}.
The authors perform both closed-set and open-set tests and show how their combination provides an accuracy gain in the considered scenarios.
The work of~\cite{wang2011channel} aims to secure \gls{asv} systems against playback attacks, which claim the victims' identity by playing back their voice recorded without consent.
The method leverages the noise pattern of the audio channel considering noises from intermediate recording and playback devices in the authentic recordings.

The second group of \gls{df} detectors relies on more semantically meaningful features and exploits high-level inconsistencies to discriminate \glspl{df}, assuming their weakness in emulating the finest aspects of human nature.
As an example, \cite{li2018ictu} focuses on the detection of the lack of natural eye blinking in synthesized videos, whereas~\cite{chugh2020not} looks for semantic mismatches between the audio and video modalities, such as the absence of lip-syncing.
The authors of~\cite{yang2019exposing} perform face-swap \gls{df} detection by comparing two different estimates of the subject's head pose and detect a \gls{df} whenever there is an apparent difference between them, indicated by a mismatch of landmark locations.
Similarly,~\cite{cozzolino2021id} detects fake videos by modeling how people move as they speak, while~\cite{agarwal2020detecting} addresses the same problem combining static and temporal bio-metrics based on face recognition and expressions/head movements.
The authors of~\cite{conti2022deepfake} show how synthetic voices lack natural emotional behavior and can be discriminated by feeding a classifier with high-level features obtained from a \gls{ser} system. 
Finally, the work presented in~\cite{hosler2021deepfakes} encompasses the two previous approaches by exploiting audio and visual emotion analysis to detect joint audio-visual \glspl{df}. 

In this paper, we adopt a semantic approach to perform \gls{df} speech detection.
We partially take inspiration from the work presented in~\cite{agarwal2020detecting}, where face-swap \glspl{df} are identified by looking at the mismatch between facial recognition static cues and behavioral bio-metrics based on expression and head movement.
Our scenario considers speaker identification aspects together with speech prosody, defined as all the information present in a speech signal but not specified in the text (e.g., temporal variations in rhythm, intonation, stress, style, etc.).
This constitutes a basis we can leverage to identify \gls{df} speech generated via different technologies that may be flawed in one semantic aspect or the other.
In particular, we represent the identity of the speaker extracting a set of embedding through a recent \gls{asv} network~\cite{desplanques2020ecapa}. At the same time, we obtain the bio-metric behavior that corresponds to the speech prosody through the use of an encoder network originally proposed for speech synthesis~\cite{skerry2018towards}. Differently from~\cite{agarwal2020detecting}, we do not define a reference set for the identity-behavioral mapping, but we feed the concatenation of speaker and prosody embeddings to a simple supervised classifier, adopting an approach similar to that proposed in~\cite{conti2022deepfake}. 
We believe that combining two semantic representations as speaker-identity and prosody can model both the voice's physiological and behavioral characteristics.  
Our detector proves capable of detecting synthetic speech samples generated with both \gls{tts} and \gls{vc} techniques.
Furthermore, our detector shows good generalization properties when we test it on unseen datasets or MP3-compressed recordings.
\section{Proposed System}
\label{sec:method}

In this work, we propose a method for synthetic speech detection named \textit{ProsoSpeaker}. This predicts if a speech recording is authentic or has been synthetically generated by analyzing the audio signal only.

\begin{figure}[t]
    \centering
    \includegraphics[width=.6\columnwidth]{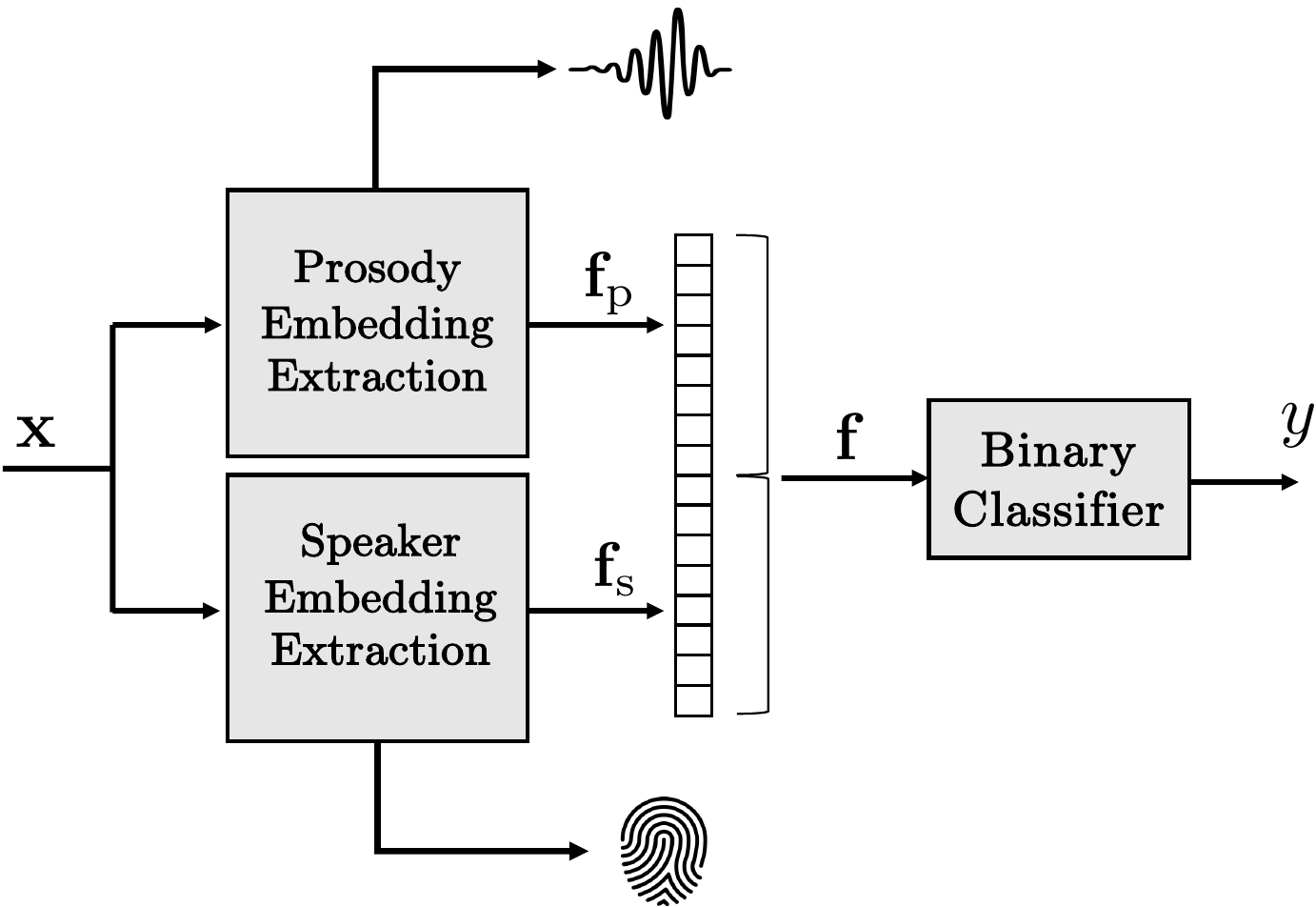}
    \caption{Pipeline of the proposed \textit{ProsoSpeaker} system.}
    \label{fig:block_scheme}
    \vspace{-1.5em}
\end{figure}

Formally, given a discrete-time input speech signal $\mathbf{x}$ sampled with sampling frequency $F_\text{s}$, the goal is to predict the associated label $y$ such that 
\begin{equation}
    y \in\{ \text{REAL}, \text{DF}\},
\end{equation}
where $\text{REAL}$ identifies authentic speech samples, while $\text{DF}$ corresponds to speech that has been synthetically generated, either using \gls{tts} or \gls{vc} technique. Figure~\ref{fig:block_scheme} shows the pipeline of the proposed system. 
As mentioned, our approach leverages the difficulty of \glspl{df} in generating complex semantic aspects of voice naturally. Hence, the proposed \textit{ProsoSpeaker} method relies on a rich set of high-level features extracted from the input $x$ and obtained as the concatenation of two embedding vectors extracted from two different networks architectures. 
We will refer to each one of them as \textit{speaker} ($\mathbf{f}_\text{s}$) and \textit{prosody} ($\mathbf{f}_\text{p}$) embeddings.
This representation is then used as input to a simple supervised classifier, which outputs for each input $\mathbf{x}$ a prediction of the label $y$.
In the following we provide additional details about each step of the pipeline depicted in Figure~\ref{fig:block_scheme}.

\subsection{Speaker Embedding Extraction}
\label{subsec:speaker_embedding}

The principle of \gls{vc} algorithms is to operate on pristine speech signals and modify their frequency content to match a target identity.
We believe that this kind of forgeries could leave traces in the speaker timbre quality that we can leverage to perform synthetic speech detection. 
We propose to do so through a feature set that describes each voice's unique fingerprint in a compact fashion, extracting the spectro-temporal characteristics of the analyzed spokesperson, i.e., timbre specific properties or pitch contour of the voice. This feature set, that we indicate with $\mathbf{f}_\text{s}$, is extracted exploiting a state-of-the-art network \cite{desplanques2020ecapa} originally proposed for a speaker recognition task.
The proposed speaker embeddings can spot voice anomalies and allow us to discriminate between real and synthetic tracks generated through \gls{vc} engines, as we will prove in the results section.

As mentioned, speaker embeddings are defined and computed using the ECAPA-\gls{tdnn} model, firstly proposed in~\cite{desplanques2020ecapa} for \gls{asv}.
This model, that takes as input the \glspl{mfcc} of the input signal, enhances the typical X-vectors architectures~\cite{snyder2018x,snyder2019speaker,zeinali2019but} and outperforms state-of-the-art \gls{tdnn} based systems.
The architecture is inspired by the latest trends in face verification and computer vision as it includes residual blocks~\cite{he2016deep} to skip connections and exploit multi-layer information, and squeeze-excitation blocks~\cite{hu2018squeeze} that explicitly model channel inter-dependencies.
In addition, the model extends the temporal attention mechanism presented in~\cite{okabe2018attentive} to be channel-dependent and better adapt to global attributes, e.g., noise or recording conditions.
This design allows the network to achieve higher generalization ability, capture high-level properties, and improve performance while significantly reducing the number of model parameters.
We use this model as an embedding extractor by first training it for the original task (i.e., \gls{asv}) and then feeding the considered input $\mathbf{x}$ to the trained network. We then assume as embedding representation the output of the network discarding the final classification layer, hence adopting a transfer-learning strategy. 
In particular, the variable length signal $\mathbf{x}$ is first pre-processed, i.e.,  transformed in time-frequency domain applying a \gls{stft} with window length $W_\text{s}$ and hop size $H_\text{s}$. From the resulting spectrogram $\mathbf{X}$ we compute a set of \glspl{mfcc}, i.e., 
\begin{equation}
    \mathbf{X}_\text{MFCC} = \text{MFCC}(\mathbf{x}) \in \mathbb{R}^{M \times B},
\end{equation}
where $M$ corresponds to the number of time windows and $B$ is the total number of mel-frequency cepstrum coefficients.
This feature map is then used as input to the trained ECAPA-\gls{tdnn} network, which projects it into the fixed-length speaker embedding  $\mathbf{f}_\text{s}$ of dimension $N_\text{s}$. 

\subsection{Prosody Embedding Extraction}
\label{subsec:prosody_embedding}

Complementary to the aspects described by the speaker embeddings, we believe that high-level prosodic aspects, like speech signal variations in rhythm, intonation and style, constitute another aspect we can leverage to discriminate deepfake speech tracks.
In particular, prosody measures an intrinsic human voice characteristic that we assume \gls{tts} synthesis algorithms struggle at recreating. 
In fact, despite the recent advances, synthetic prosody has different quality and intensity w.r.t. to human speech, and this difference can be captured using a set of prosody embeddings.
This assumption is later proved by the obtained results.
The prosody embedding vector $\mathbf{f}_\text{p}$ we propose corresponds to the result of the reference encoder of the model presented in~\cite{skerry2018towards}, which we will refer to as prosody encoder.

The prosody encoder \cite{skerry2018towards} was initially introduced to improve the naturalness of the voices synthesized by Tacotron~\cite{wang2017tacotron,wang2018style} enhancing their prosody controls.
The whole Tacotron model receives a text as input and generates speech depending on the speaker's identity considered in the training phase. 
In contrast, the encoder takes as input the mel-spectrogram transform of a reference signal conveying the desired prosody and extracts a fixed-length learned representation.
This is used to condition the synthesis, making it possibly more expressive.
The authors show that the results match the prosody with fine temporal detail even when the target and reference speakers are different. 
The prosody encoder comprises a 6-layer stack of 2D convolutions with batch normalization, followed by a \gls{gru} layer to summarize the variable-length sequence. 
Finally, a fully-connected layer extracts the embeddings in the desired dimension.
This design sufficiently bottlenecks the input information such that the encoder is forced to learn a compact representation of prosody. 
For this work, we train Tacotron and prosody encoder jointly by synthesizing target audio signals, provided as input to both, and using the reconstruction error as loss function. Once the prosody encoder is trained, we use it as an embedding extractor, feeding as input the mel-spectrogram of the input signal $\mathbf{x}$
\begin{equation}
    \mathbf{X}_\text{mel} = \text{MelSpec}(\mathbf{x}) \in \mathbb{R}^{M \times K},
\end{equation}
where $M$ is the number of time windows and $K$ corresponds to the total number of frequency bins, extracted with window size $W_\text{p}$ and hop size $H_\text{p}$. The output of the prosody encoder is the vector  $\mathbf{f}_\text{p}$ of length $N_\text{p}$.

\subsection{Binary Classifier}
As shown in Figure~\ref{fig:block_scheme}, the final part of the \textit{ProsoSpeaker} pipeline is a supervised binary classifier. We concatenate the two embeddings $\mathbf{f}_\text{s}$ and $\mathbf{f}_\text{p}$ obtaining a final feature vector
\begin{equation}
    \mathbf{f} = [\mathbf{f}_\text{s}, \mathbf{f}_\text{p}] \in \mathbb{R}^{N_\text{s} + N_\text{p}},
\end{equation}
which is fed to the classification stage.
The supervised classifier is trained to predict the class $y$ of the input speech $\mathbf{x}$. We decide to adopt a simple classification front-end because we mostly rely on the discriminative capacity of the rich proposed feature set. Moreover, it is worth noting that any supervised classifier algorithm can be used at this stage, as our pipeline is classifier-independent.

\section{Experimental Setup}
\label{sec:setup}
In this section we provide the reader some insights on the evaluation setup used to assess the performances of the \textit{ProsoSpeaker} detector. We first describe the dataset used for training and testing the system. Then, we specify all the training parameters for both the back-end ( i.e., the embedding extractors) and front-end (i.e., the binary supervised classifier). Finally, we describe the training process and the selected baselines.

\subsection{Dataset Description}
\label{subsec:dataset}

In this section we introduce the datasets involved in the training and testing phases of the presented work, which in total counts almost 800000 tracks. 
We considered multiple datasets containing tracks of both REAL (i.e, authentic) and DF (i.e., synthetic) classes, aiming to test the proposed method's generalization properties.
We set the sampling frequency $F_\text{s}$ to $16$~kHz during all the experiments, hence if necessary, down-sampling the audio tracks.
In the following we provide further details for each dataset that will be later useful in interpreting the experimental results.

\vspace{0.3em}\noindent\textbf{ASVspoof 2019}~\cite{todisco2019asvspoof}
is a speech audio dataset containing both real and synthetic tracks. It has been released for the ASVspoof challenge, in which participants compete to implement the best anti-spoofing system for \gls{asv}. 
Here we consider the \gls{la} partition of the dataset, further divided in \textit{train}, \textit{dev} and \textit{eval}, which includes spoofing atalzantot2019deeptacks generated through \gls{tts}, \gls{vc} and \gls{tts}/\gls{vc} hybrid techniques.
Each partition comprises authentic signals along with speech samples generated with 19 different synthesis algorithms. The \textit{train} and \textit{dev} partitions have been created using the same set of synthesis algorithms (named $A01$, $A02$, ..., $A06$), while the \textit{eval} partition includes samples generated with different techniques ($A07$, ..., $A19$). We use \textit{train} and \textit{dev} partitions for training and fine-tuning the proposed method, while \textit{eval} partition is used in test. 

An updated version of the dataset was released in 2021 (\cite{yamagishi2021asvspoof}). Nevertheless, we decided not to consider it since, at the time of writing, it is distributed only with REAL/DF labels, while no information is available about the generation strategy adopted for each audio track.

\vspace{0.3em}\noindent\textbf{LibriSpeech}~\cite{panayotov2015librispeech} is a dataset containing about 1000 hours of authentic speech from different speakers. From this corpus we considered the subset \textit{train-clean-100}. We include audio tracks from this dataset in the training set.
    
\vspace{0.3em}\noindent\textbf{\gls{ljs}}~\cite{LJSpeech} is a dataset containing short audio tracks of REAL speech recorded from a single speaker reciting pieces from non-fiction books. This dataset is part of the test set.
    
\vspace{0.3em}\noindent\textbf{Cloud2019} is a collection of \gls{tts} generated audio signals proposed in~\cite{lieto2019hello}.
It includes tracks from different speech generators available as cloud services: \gls{po}, \gls{gs}, \gls{gw}, \gls{az} and \gls{wa}. 
We include this dataset in the test set as \gls{df} signals.
    
\vspace{0.3em}\noindent\textbf{\gls{iemocap}(IEM)}~\cite{busso2008iemocap} is a dataset originally designed for the \gls{ser} task. 
The data were recorded during scripted and improvised conversations by 10 actors.
It contains video and audio signals annotated with information about the speakers' facial expressions and head movements.
We include this dataset in the test set as authentic signals.

\vspace{0.3em}
Table~\ref{tab:datasets} reports the train and test split used for the front-end binary classifier and the type of speech signal, REAL or DF, included in each dataset.

\newcolumntype{Y}{>{\centering\arraybackslash}X}

\begin{table}
\small
\centering
\caption{Composition of the training, development and test sets for the proposed experiments.}
\label{tab:datasets}
\begin{tabularx}{\textwidth}{l|Y|YY|YYY}
\toprule
\textbf{Dataset} & \textbf{N.~Tracks} & \textbf{REAL} & \textbf{DF}  & \textbf{Train} & \textbf{Dev} & \textbf{Test} \\ \midrule \midrule
ASVspoof 2019 & 121\,458       & \text{\ding{55}}  & \text{\ding{55}} & \text{\ding{55}} & \text{\ding{55}}   & \text{\ding{55}}  \\
LibriSpeech   & 28\,539       & \text{\ding{55}}  &     & \text{\ding{55}}   &     & \\
LJSpeech      & 13\,100       & \text{\ding{55}}  &     &    &       &\text{\ding{55}}  \\
Cloud2019     & 11\,888       &      & \text{\ding{55}} &       &     & \text{\ding{55}}  \\
IEMOCAP       & 10\,039       & \text{\ding{55}}  &     &       &   & \text{\ding{55}} \\ 
\midrule
\textbf{Total} & 185\,024 & 64\,159 & 120\,865 & 53\,919 & 24\,844 & 106\,264 \\
\bottomrule
\end{tabularx}

\vspace{-2em}
\end{table}

\subsection{Training}
Our system involves the training of three independent blocks: the ECAPA-\gls{tdnn} network, the prosody encoder, the final binary classifier. 
Regarding the speaker embedding extractor, we use a version of ECAPA-\gls{tdnn} available at~\cite{ravanelli2021speechbrain}, which uses Additive Margin Softmax Loss and is trained on VoxCeleb 1~\cite{nagrani2017voxceleb} and VoxCeleb 2~\cite{chung2018voxceleb2} datasets.
As mentioned in Section~\ref{subsec:speaker_embedding}, the input waveform $\mathbf{x}$, to be used as input to ECAPA-\gls{tdnn} network, must be first transformed in its \gls{mfcc} representation $\mathbf{X}_\text{MFCC}$. For this operation we consider $B = 80$ \glspl{mfcc} extracted with $W_\text{s} = 25$~ms windows with hop size $H_\text{s} = 10$~ms, leading to a $M \times 80$ representation, where the number of windows $M$ depends on the length of the audio. The final embedding vector $\mathbf{f}_\text{s}$ has dimension $N_\text{s} = 192$.
For the prosody embedding extractor, we train the prosody encoder on Blizzard 2013 dataset~\cite{king2013}, following the training procedure detailed in~\cite{skerry2018towards}. For computational issues, we modify only one parameter value, the mini-batch size, that in our training process is equal to 8.
Before feeding it to the encoder, the input signal $\mathbf{x}$ is transformed into a mel-spectrogram $\mathbf{X}_\text{mel}$ using window length $W_\text{p} = 50$~ms and hop size $H_\text{p} = 12.5$~ms. The number of frequency bins used is $K = 80$. This lead to a final input of dimension $M \times 80$.
The resulting embedding vector $\mathbf{f}_\text{p}$ has length and $N_\text{p} = 128$.
The final concatenated feature set is $\mathbf{f}$ of length $N =N_\text{s}+N_\text{p}=320$.
This vector is standardized using z-score, i.e., removing the mean and scaling to unit variance, and acts as input to the binary classifier. 
The supervised classification algorithm we adopt is \gls{svm} classifier, following the training-development partition detailed in Table~\ref{tab:datasets}.
To find the best set of hyper-parameters we performed a grid search on development partition using balanced accuracy as a metric. We considered the following parameters: $C \in[0.01, 0.1, 1, 10, 100]$, kernel coefficient $\gamma \in [1/N, 1 / (N * \sigma^2_\mathbf{f})]$, where $N$ is dimensionality of the feature vector $\mathbf{f}$ and $\sigma^2_\mathbf{f}$ is the variance of $\mathbf{f}$ over the training dataset. In addition, we vary the kernel type between radial basis function kernel, polynomial kernel and sigmoid kernel. The best configuration proved to be $C = 100$, $\gamma = 1 / (N * \sigma^2_\mathbf{f})$ and using radial basis function kernel.

\subsection{Baselines}
To test the validity of our method, we compare its performances with those of three different baselines.
The first one is RawNet2~\cite{tak2021end}, a state-of-the-art end-to-end neural network that operates on raw waveforms.
It has been first proposed for the ASVspoof 2019 challenge and included as a baseline in the ASVspoof 2021 challenge both for \gls{la} and \gls{df} tasks.
The second and third baselines are variants of ResNet \cite{he2016deep}, a residual Convolutional Neural Network that creates shortcuts between layers by skipping connections that help stabilize training.
We consider two versions of the ResNet, fed with different representations of the input audio track, as presented in~\cite{alzantot2019deep}.
The first one, referenced as Spec-ResNet, takes as input the log-magnitude representation of the \gls{stft} of the considered audio.
The second one, called MFCC-ResNet, is fed with the \glspl{mfcc} of the input data, together with their first and second derivatives. Input transformation and training strategy for these two networks are implemented following~\cite{alzantot2019deep}.
The three baselines considered are based on different representations of the input data, allowing us to have an orthogonal approach to the problem.
All the models share the same training set we adopted for the proposed method.
\section{Results}
\label{sec:results}

In this section we assess the performances of \textit{ProsoSpeaker} detector, measuring the performances of the method in terms of \gls{roc} curves, \gls{auc}, \gls{eer} and balanced accuracy.
Optimal performances are reached when \gls{auc} and balanced accuracy are equal to one, while \gls{eer} is equal to 0.
All the models presented in the following have been trained on the same dataset, obtained by the union of ASVspoof 2019 \gls{la} and LibriSpeech, as shown in Table~\ref{tab:datasets}. 

\subsection{Baseline comparison}
\label{subsec:baseline}

\begin{figure}[t]
    \centering
    \includegraphics[width=.5\columnwidth]{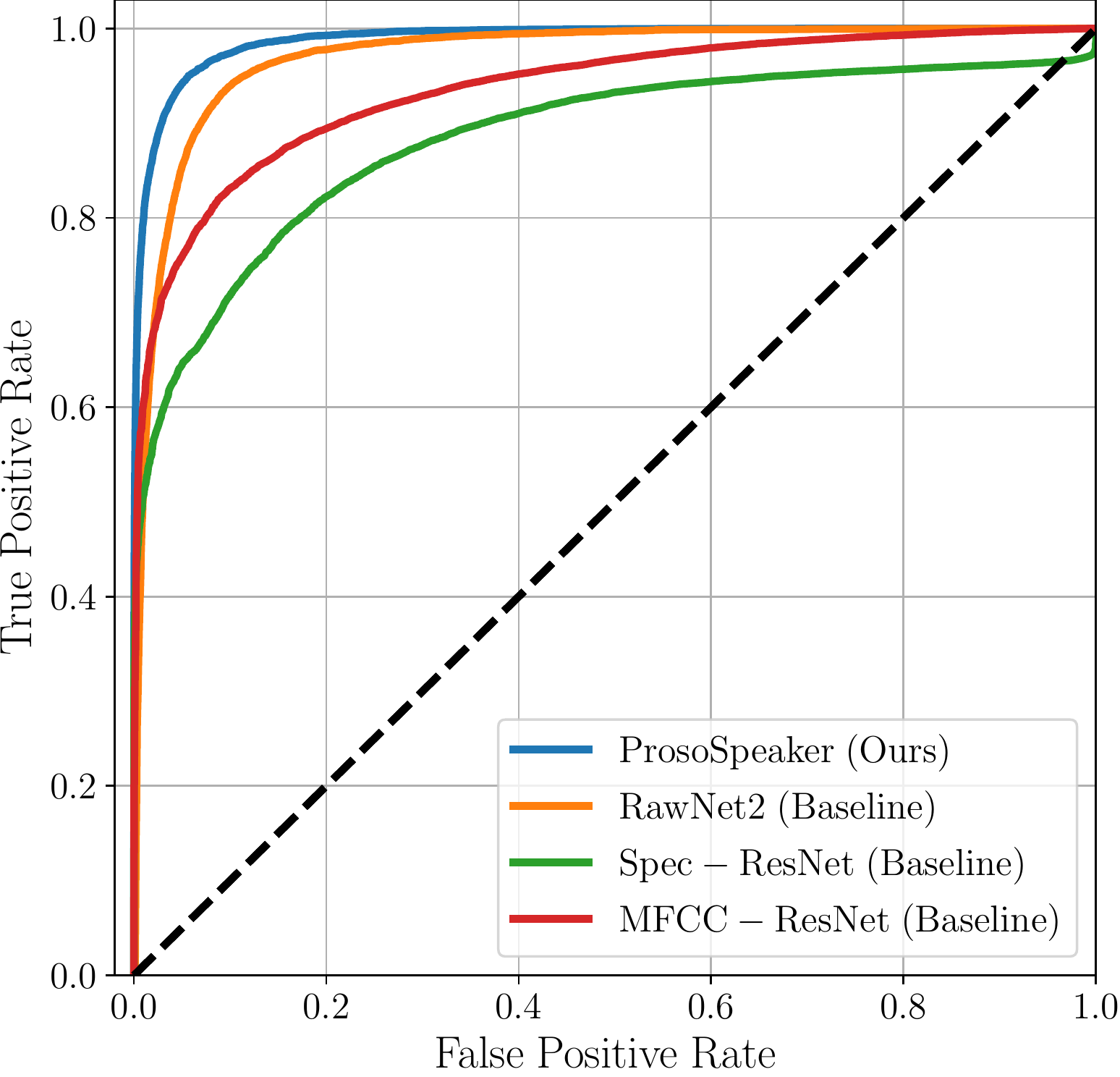}
    \caption{\gls{roc} curves for the proposed method and the considered baselines, evaluated on ASVspoof 2019 \gls{la} \textit{eval} set.}
    \label{fig:asv2019}
    \vspace{-1.5em}
\end{figure}

As a first experiment, we compare the results obtained using the proposed method with those of the considered baselines on the \gls{la} \textit{eval} partition of the ASVspoof 2019 dataset.
Figure~\ref{fig:asv2019} shows the \gls{roc} curves of the three detectors and Table~\ref{tab:baselines_2019} shows the corresponding \gls{auc}, \gls{eer} and balanced accuracy values.
\textit{ProsoSpeaker} detector outperforms all the three baselines in the considered metrics.
In particular, the most remarkable improvement is seen over Spec-ResNet with a difference of about $15\%$ for \gls{eer} and balanced accuracy, $10$ for \gls{auc}.
A significant gain is also shown concerning RawNet2.
Here, our method improves by almost $3\%$ over \gls{eer} and balanced accuracy, while $2$ over \gls{auc}.

\begin{table}
\small
\centering
\caption{\gls{eer}, \gls{auc} and balanced accuracy values for the proposed \textit{ProsoSpeaker} method and the considered baselines, evaluated on ASVspoof 2019 \gls{la} \textit{eval} set.}
\label{tab:baselines_2019}
\begin{tabular}{@{}cccc@{}}
\toprule
\textbf{Model} & \textbf{EER \%} & \textbf{AUC} & \textbf{Bal. Acc. \%} \\ \midrule
\midrule
{RawNet2 (Baseline)} & 8.15 & 97.09 & 91.66 \\
{MFCC-ResNet (Baseline)} & 13.98 & 93.52 & 84.96 \\
{Spec-ResNet (Baseline)} & 18.75 & 88.31 & 79.50 \\ 
{\textit{ProsoSpeaker} (Ours)} & \textbf{5.39} & \textbf{98.85} & \textbf{94.43} \\
 \bottomrule
\end{tabular}
\end{table}

\subsection{Embedding analysis and ablation study}
\label{subsec:ablation}

In this second experiment we further analyze the characteristics and the importance of each embedding subset, namely the prosody embeddings $\mathbf{f}_\text{p}$ and the speaker embeddings $\mathbf{f}_\text{s}$, used in \textit{ProsoSpeaker} method.

\begin{figure}[t]
    \centering
    \includegraphics[width=0.5\columnwidth]{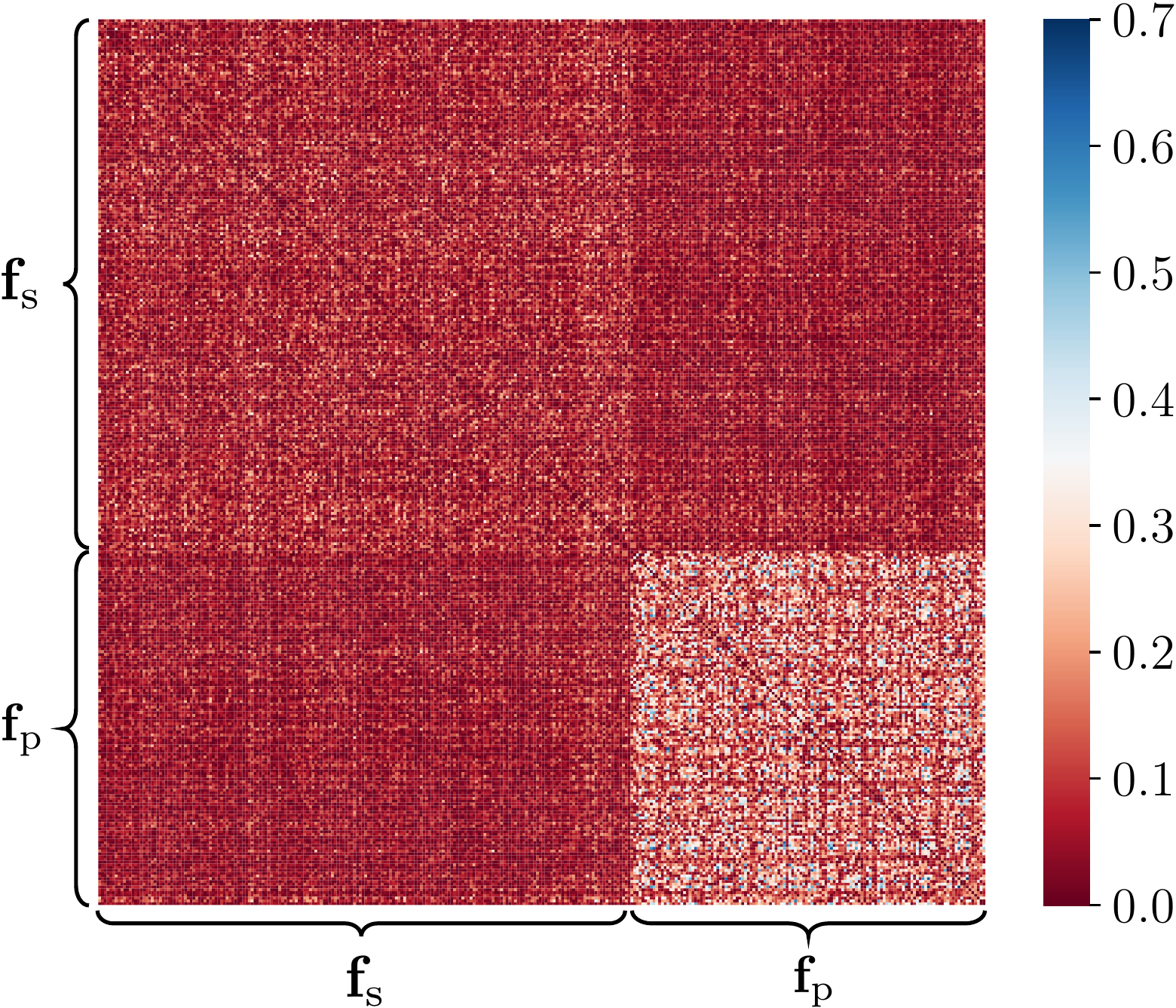}
    \caption{Cross-correlation matrix $ \mathbf{R}_{\mathbf{f} \mathbf{f}}$ of feature vectors $\mathbf{f}$ realizations of ASVspoof 2019 \textit{eval} set.}
    \label{fig:corr_eval}
\end{figure}

The first question may be how much speaker and prosody embeddings differ from each other to avoid the computation of redundant information.
To do so, we measure the sample Pearson correlation coefficient $r_{f_i f_j}$ for each pair of elements $(f_i, f_j)$ of the vector $\mathbf{f} = [f_0, f_1, ..., f_{N-1}]$ over the test dataset. The resulting matrix $ \mathbf{R}_{\mathbf{f} \mathbf{f}}$ describes both cross-correlation between prosody and speaker embeddings $\mathbf{R}_{\mathbf{f}_\text{s} \mathbf{f}_\text{p}} = \mathbf{R}_{\mathbf{f}_\text{p} \mathbf{f}_\text{s}}^T$ both auto-correlations of each embedding vector $\mathbf{R}_{\mathbf{f}_\text{p} \mathbf{f}_\text{p}}$ and $\mathbf{R}_{\mathbf{f}_\text{s} \mathbf{f}_\text{s}}$.
Figure~\ref{fig:corr_eval} shows the results of this analysis computed in the ASVspoof 2019 \textit{eval} partition. The diagonal has been set to 0 for visualization purposes.
There, we can identify two rectangular regions, one at the top left, corresponding to $\mathbf{R}_{\mathbf{f}_\text{s} \mathbf{f}_\text{s}}$, and one at the bottom right, corresponding to $\mathbf{R}_{\mathbf{f}_\text{p} \mathbf{f}_\text{p}}$. 
Although the elements of $\mathbf{f}_\text{p}$ have a higher degree of internal correlation than those of $\mathbf{f}_\text{s}$, with mean value $\mu(\mathbf{R}_{\mathbf{f}_\text{p} \mathbf{f}_\text{p}}) = 0.21$ and standard deviation $\sigma(\mathbf{R}_{\mathbf{f}_\text{p} \mathbf{f}_\text{p}}) = 0.08$, respectively, the cross coefficients present low values, with an average value of $\mu(\mathbf{R}_{\mathbf{f}_\text{s} \mathbf{f}_\text{p}}) = 0.07$.
This means that the two embedding vectors do not strongly correlate with each other and do not share much information.
The spectro-temporal and prosodic characteristics we are considering have turned out to be orthogonal to each other, benefiting our detector.

Given these results, we test how the embedding types perform individually in different scenarios.
In this analysis, we consider three distinct models, all based on the proposed architecture, differing only for the embeddings subset that the final \gls{svm} classifier receives as input.
The first model, that we indicate with \textit{Prosody Emb}, is fully-prosodic and based on $\mathbf{f}_\text{p}$ only. The second only considers the speaker information of $\mathbf{f}_\text{s}$ and we indicate it as \textit{Speaker Emb}. The third model is the complete one, i.e., \textit{ProsoSpeaker}, and it performs classification using the concatenation of $\mathbf{f}_\text{p}$ and $\mathbf{f}_\text{s}$. All three models are trained on the same dataset, i.e., ASVspoof 2019 + LibriSpeech, with the same parameters.
We then considered three test scenarios, depending on the synthesis techniques used to generate the synthetic speech signals of the test set. In the first scenario (a) we consider only speech tracks created with \gls{tts} techniques; in the second scenario (b) only speech tracks created with \gls{vc} techniques; in the third scenario (c) both synthesis techniques are considered. All the tracks for the three scenarios are selected from ASVspoof 2019 dataset.
Table~\ref{tab:ablation_table} and Figure~\ref{fig:ROC_curves} show the binary classification performances of this analysis, the first in terms of \gls{eer}, \gls{auc} and balance accuracy obtained for the three models in the three test scenarios, the second showing the corresponding \gls{roc} curves.
\begin{table}[t]
\small
\centering
\caption{\gls{eer}, \gls{auc} and Balanced Accuracy values for the three models (\textit{ProsoSpeaker}, \textit{Speaker Emb}, \textit{Prosody Emb}) tested on the three scenarios (TTS, VC, ALL).}
\label{tab:ablation_table}
\resizebox{\columnwidth}{!}{
\begin{tabular}{@{}c@{\hskip .15in}c@{\hskip .1in}c@{\hskip .1in}c@{\hskip .15in}c@{\hskip .1in}c@{\hskip .1in}c@{\hskip .15in}c@{\hskip .1in}c@{\hskip .1in}c@{}}
\toprule
\multicolumn{1}{l}{} & \multicolumn{3}{c}{\textbf{(a) TTS}} & \multicolumn{3}{c}{\textbf{(b) VC}} & \multicolumn{3}{c}{\textbf{(c) ALL}} \\ \midrule
\multicolumn{1}{l}{} & \textbf{EER} & \textbf{AUC} & \textbf{BA} & \textbf{EER} & \textbf{AUC} & \textbf{BA} & \textbf{EER} & \textbf{AUC} & \textbf{BA} \\ \midrule
\textbf{\textit{ProsoSpeaker}} & \textbf{4.93} & \textbf{99.02} & \textbf{94.77} &   \textbf{6.70} & \textbf{98.29} & \textbf{93.28} & \textbf{5.39} &   \textbf{98.85} & \textbf{94.43} \\
\textbf{\textit{Prosody Emb}} & 8.58 & 96.68 & 90.64 & 30.04 & 76.35 & 66.44 & 15.13 & 91.99 & 85.05 \\
\textbf{\textit{Speaker Emb}} & 26.21 & 81.64 & 74.62 & 9.82 & 96.53 & 88.20 & 22.88 & 85.08 & 77.75 \\ \bottomrule
\end{tabular}
}
\end{table}

\newcounter{mysfig}
\counterwithin{mysfig}{figure}

\renewcommand\themysfig{(\alph{mysfig})}
\makeatletter
\newcommand\Scaption[1]{%
\refstepcounter{mysfig}%
\vskip.5\abovecaptionskip
  \sbox\@tempboxa{\small\themysfig~#1}%
  \ifdim \wd\@tempboxa >\hsize
    \small\themysfig~#1\par
  \else
    \global \@minipagefalse
    \hb@xt@\hsize{\hfil\box\@tempboxa\hfil}%
  \fi
  \vskip\belowcaptionskip}
\makeatother

\begin{figure}[t]
\minipage{0.32\textwidth}
  \includegraphics[width=\linewidth]{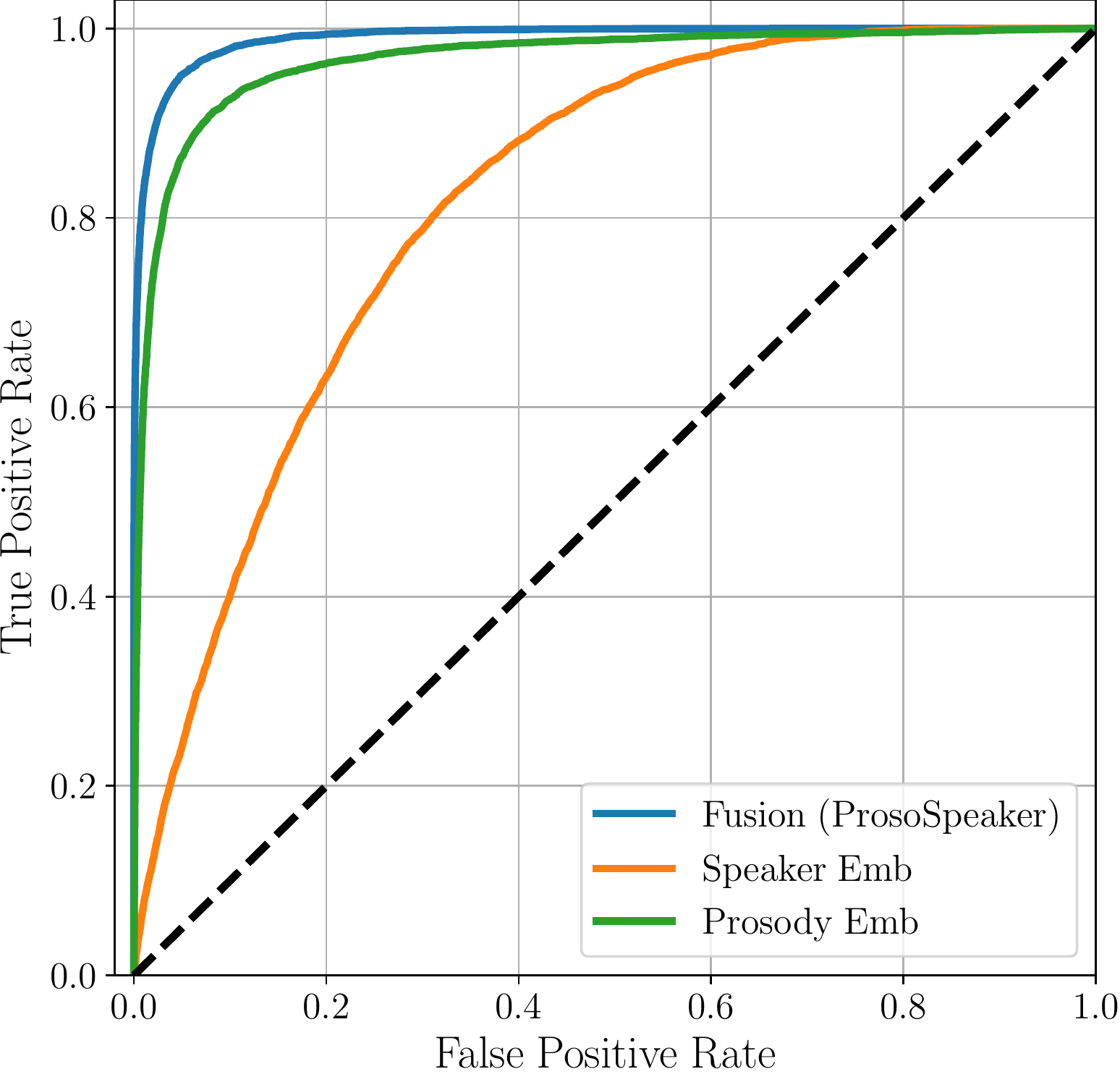}
  \Scaption{TTS}\label{fig:ROC_TTS}
\endminipage\hfill
\minipage{0.32\textwidth}
  \includegraphics[width=\linewidth]{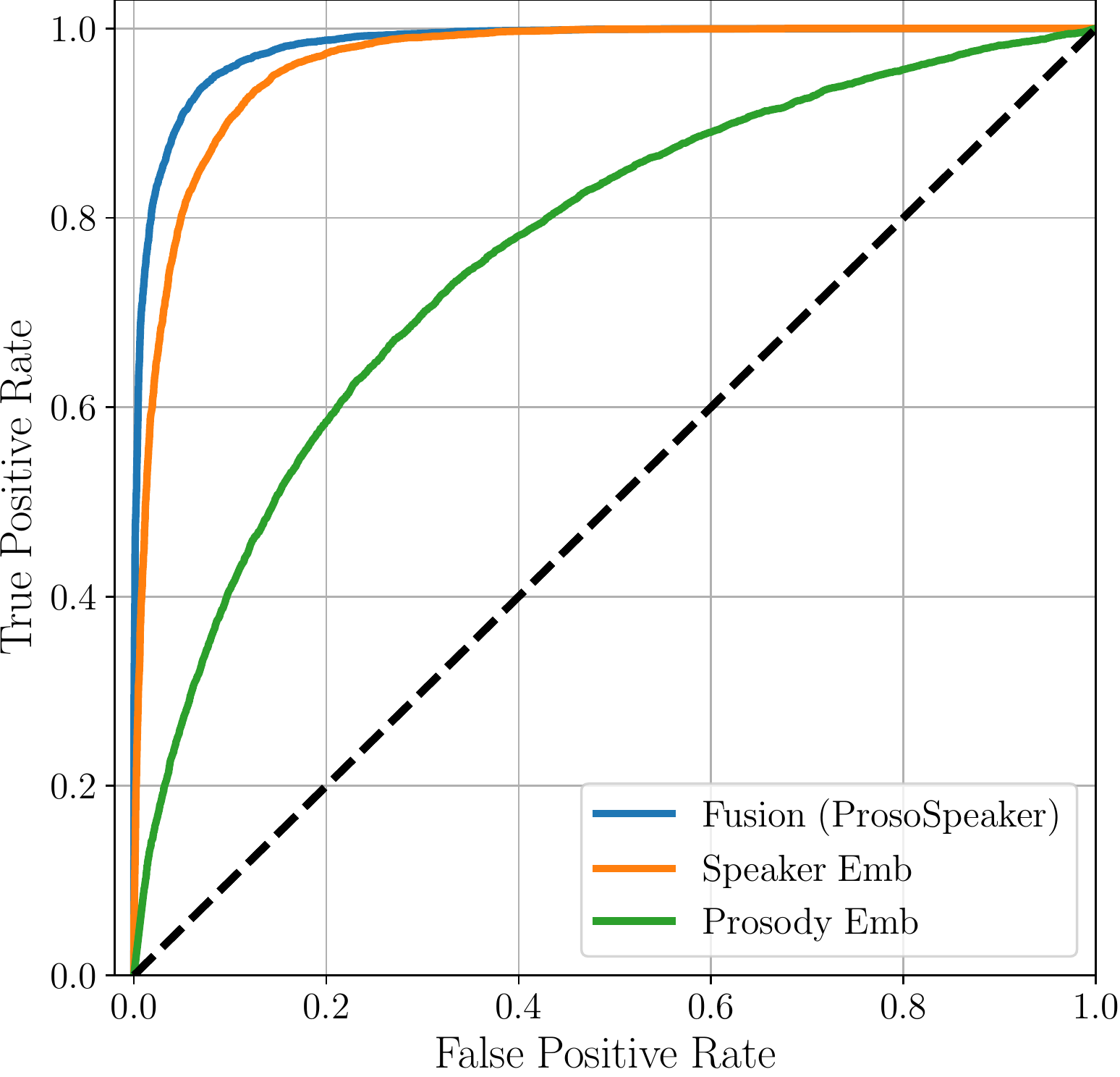}
  \Scaption{VC}\label{fig:ROC_VC}
\endminipage\hfill
\minipage{0.32\textwidth}%
  \includegraphics[width=\linewidth]{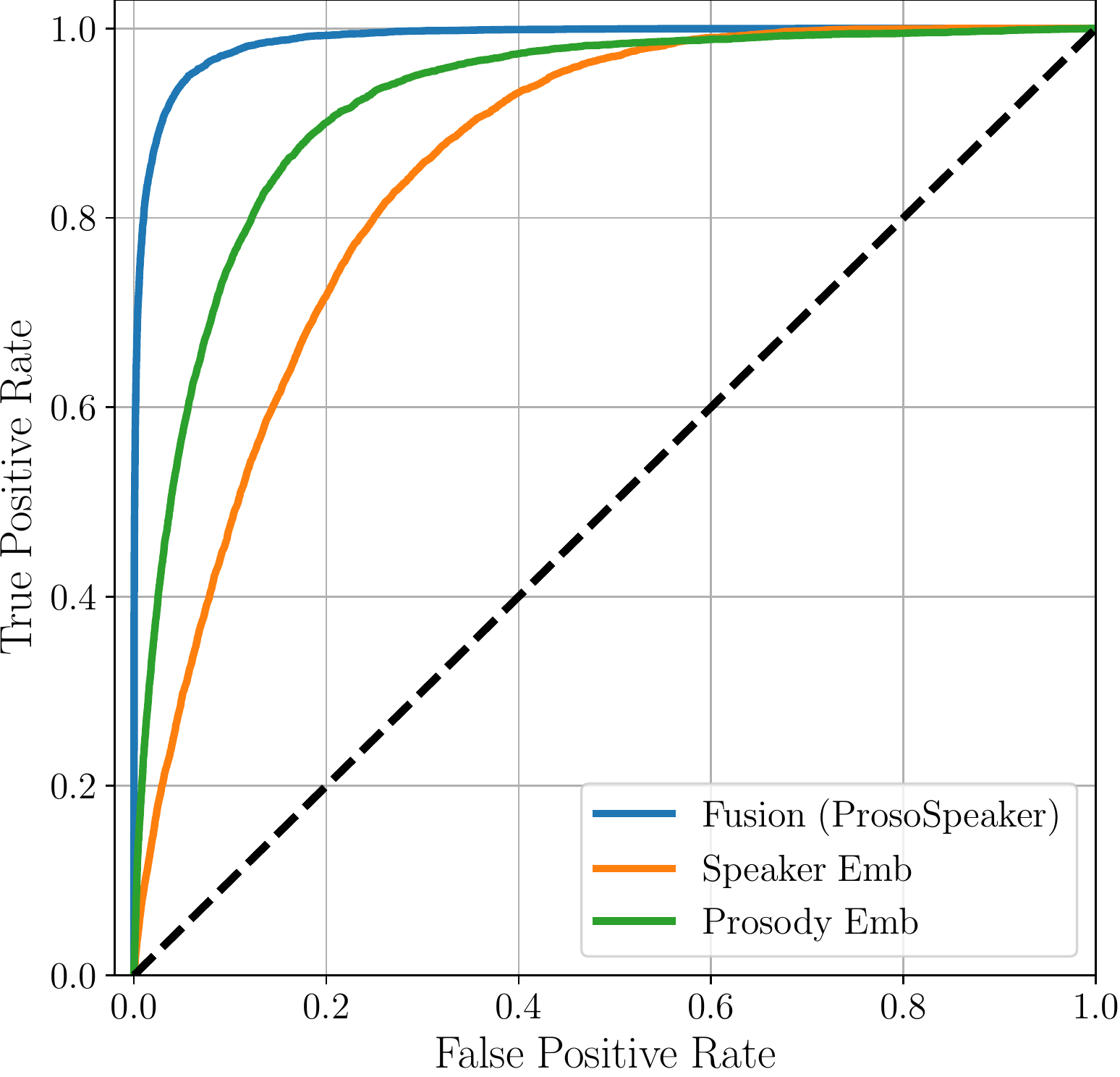}
  \Scaption{ALL}\label{fig:ROC_ALL}
\endminipage
\caption{\gls{roc} curves obtained for the three models using different embeddings (\textit{ProsoSpeaker}, \textit{Speaker Emb}, \textit{Prosody Emb}) and tested on the three scenarios (TTS, VC, ALL).}
\label{fig:ROC_curves}
\end{figure}

The predictions of the two partial models are orthogonal to each other and each performs better on a distinct scenario.
In particular, prosodic embeddings $\mathbf{f}_\text{p}$ can discriminate speech signals generated with \gls{tts} algorithms well but are less effective with \gls{vc} methods, while speaker embeddings $\mathbf{f}_\text{s}$ achieves better results in the \gls{vc} case than \gls{tts}.
From these results we can confirm our initial hypothesis, i.e., that each one of the two speech generation techniques fails in reproducing one of the semantic features encoded by $\mathbf{f}_\text{s}$ or $\mathbf{f}_\text{p}$.
On one side, \gls{tts} systems struggle at recreating natural sounding prosody, starting from a pure textual input, and hence prosody embeddings are effectively discriminating them from real speech. 
On the other hand, \gls{vc} techniques manipulate an authentic speech sample to impersonate a target speaker, introducing artifacts in the timbre qualities that can be detected leveraging speaker embeddings.
Nonetheless, the fusion of the two embeddings improves the predictions in all the considered scenarios, reaching an \gls{auc} = 0.99 in the case of the complete dataset. We can conclude that the concatenation of the two embeddings provides a more comprehensive and significant representation of the input speech signal, leading to higher binary classification performances. 

\subsection{Generalization}
\label{subsec:generalization}

\begin{figure}[t]
    \centering
    \includegraphics[width=0.85\columnwidth]{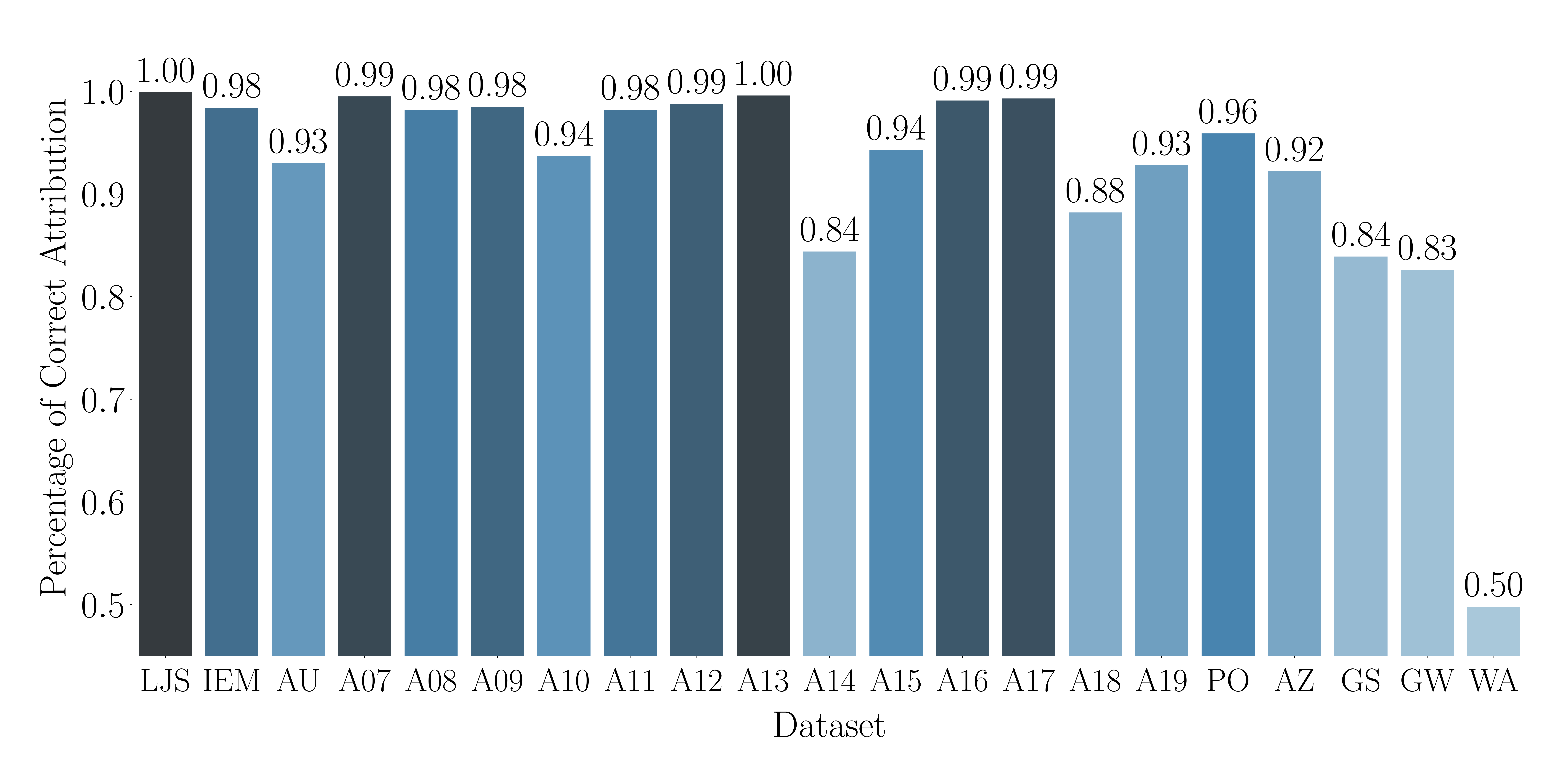}
    \caption{Bar plot of the percentage of correct attribution values of the proposed model on each partition of each considered dataset.}
    \label{fig:acc_barplot}
\end{figure}

In this third set of experiments, we aim to analyze the consistency and generalization ability of the proposed method by augmenting the considered test set.
First, we verify the performances of the proposed detector singularly on each algorithm present in ASVspoof 2019 \textit{eval} to check the classification performances consistency over different synthesis strategies. Then, we want to assess \textit{ProsoSpeaker}'s generalization capabilities across multiple datasets, unseen during training and external to the ASVspoof challenge corpora.
Figure~\ref{fig:acc_barplot} shows the percentage of correct attribution values obtained for each synthesis algorithm included in ASVspoof 2019 \textit{eval} set (A07, A08, ..., A13) and for LJSpeech, IEMOCAP and Cloud2019 (divided in PO, AZ, GS, GW, WA). The label AU corresponds to real speech samples distributed in ASVspoof 2019. 
The proposed method is successful in almost all the considered cases, with a percentage of correct attribution value always higher than 0.80.
This means that \textit{ProsoSpeaker} has good generalization capabilities, and we can consider it a reliable method.
The only exception is represented by the \gls{tts} generator IBM Watson, included in Cloud2019, where the accuracy is equal to 0.50.
We believe this issue is due to the fact that the IBM \gls{tts} method is specifically trained considering a ``prosodic-phonology'' approach for generating expressive speech~\cite{pitrelli2006ibm}, hence deceiving our detection method.

\subsection{Robustness analysis}
\label{subsec:robustness}

Some additional tests are finally necessary to verify the robustness of the proposed method to common signal manipulation, i.e., compression.
In fact, in a real-world scenario, many operations can be performed to hide the artifacts introduced by deepfake generation algorithms, like, for instance, lossy compression. Some signal information is lost by compressing an audio track, including traces that may help deepfake detectors determine the signal's authenticity.
Since our method does not rely on low-level signal characteristics but analyzes semantic features, we hypothesize that compression should not affect its performance significantly. In practice, speaker and prosody embeddings should be only partially impacted by this type of data augmentation and keep their discriminative potential.
To test such aspect, we create three versions of the ASVspoof 2019 \gls{la} \textit{eval} dataset using MP3 compression at different bitrates, namely 128 kBits/s, 64 kBits/s and 32 kBits/s, using SoX tool~\cite{soxcompression}.
Table~\ref{tab:results_compression} shows the correspondent \gls{auc}, \gls{eer} and balanced accuracy values for different compression bitrates.
The detector's performance deteriorates as we increase the compression factor, observing \gls{auc} and \gls{eer} values dropping by 2 and 4\%, respectively, between the two extreme cases. Balanced accuracy decreases significantly when compression is firstly introduced, with a drop of 4\% between the no-compression and 128 kBits/s cases. At the same time, it maintains stable values when the bitrate decreases, falling only by 1\% between 128 and 32 kBit/s cases.
We can conclude that, overall, the proposed system, thanks to its high-level semantic approach, is able to maintain its effectiveness even in presence of heavy signal compression.

\begin{table}
\small
\centering
\caption{\gls{roc} \gls{auc}, \gls{eer} and Bal. Acc. values computed on compressed versions of ASVspoof 2019 \gls{la} \textit{eval} at different bitrates.}
\label{tab:results_compression}
\begin{tabular}{@{}c@{\hskip .3in}c@{\hskip .3in}c@{\hskip .3in}c@{}}
\toprule
\textbf{Compression Rate} & \textbf{ROC AUC} & \textbf{EER \%} & \textbf{Bal. Acc. \%} \\ \midrule
\midrule
{No Compression} & 98.86 & 5.21 & 94.40 \\
{128 kBits/s} & 98.35 & 6.96 & 89.83 \\
{64 kBits/s} & 98.13 & 7.12 & 89.68 \\ 
{32 kBits/s} & 96.77 & 9.81 & 88.50 \\
 \bottomrule
\end{tabular}
\vspace{-2em}
\end{table}

\section{Conclusions}
\label{sec:conclusions}

In this paper we presented a novel method to perform \gls{df} speech detection. 
Adopting a semantic approach, we based our system on the concatenation of high-level features, denoted as speaker and prosody embeddings.
This representation is used as input to a fast supervised binary classifier that predicts whether the speech signal is authentic or synthetically generated.
We have shown that the performance of the proposed method outperforms those of the state-of-the-art considered baselines. In addition to that, it presents good generalization properties and is robust to real-world audio manipulation, as lossy compression. 
Moreover, through an ablation study, we observed how speaker and prosody embeddings perform individually in different scenarios and why their combination is the more effective strategy, achieving higher classification performances.
The obtained results validate the idea of exploiting semantic features to discriminate deepfakes and highlight some of the aspects on which speech generators still fail.

\subsubsection{Acknowledgements} This work was supported by the PREMIER project, funded by the Italian Ministry of Education, University, and Research within the PRIN 2017 program. This material is based on research sponsored by DARPA and Air Force Research Laboratory (AFRL) under agreement number FA8750-20-2-1004. The U.S. Government is authorized to reproduce and distribute reprints for Governmental purposes notwithstanding any copyright notation thereon. The views and conclusions contained herein are those of the authors and should not be interpreted as necessarily representing the official policies or endorsements, either expressed or implied, of DARPA and Air Force Research Laboratory (AFRL) or the U.S. Government.

\bibliographystyle{splncs04}
\bibliography{bibliography}

\begin{thebibliography}{10}
\providecommand{\url}[1]{\texttt{#1}}
\providecommand{\urlprefix}{URL }
\providecommand{\doi}[1]{https://doi.org/#1}

\bibitem{agarwal2021detecting}
Agarwal, S., Farid, H.: {Detecting Deep-Fake Videos From Aural and Oral
  Dynamics}. In: IEEE/CVF Conference on Computer Vision and Pattern Recognition
  (CVPR) (2021)

\bibitem{agarwal2020detecting}
Agarwal, S., Farid, H., El-Gaaly, T., Lim, S.N.: {Detecting deep-fake videos
  from appearance and behavior}. In: IEEE International Workshop on Information
  Forensics and Security (WIFS) (2020)

\bibitem{alzantot2019deep}
Alzantot, M., Wang, Z., Srivastava, M.B.: {Deep residual neural networks for
  audio spoofing detection}. In: Conference of the International Speech
  Communication Association (INTERSPEECH) (2019)

\bibitem{bonettini2020video}
Bonettini, N., Cannas, E.D., Mandelli, S., Bondi, L., Bestagini, P., Tubaro,
  S.: {Video Face Manipulation Detection Through Ensemble of CNNs}. In:
  International Conference on Pattern Recognition (ICPR) (2021)

\bibitem{borrelli2021synthetic}
Borrelli, C., Bestagini, P., Antonacci, F., Sarti, A., Tubaro, S.: Synthetic
  speech detection through short-term and long-term prediction traces. EURASIP
  Journal on Information Security  \textbf{2021}(1),  1--14 (2021)

\bibitem{busso2008iemocap}
Busso, C., Bulut, M., Lee, C.C., Kazemzadeh, A., Mower, E., Kim, S., Chang,
  J.N., Lee, S., Narayanan, S.S.: {IEMOCAP: Interactive emotional dyadic motion
  capture database}. Language resources and evaluation  \textbf{42}(4),
  335--359 (2008)

\bibitem{chen2020generalization}
Chen, T., Kumar, A., Nagarsheth, P., Sivaraman, G., Khoury, E.: Generalization
  of audio deepfake detection. In: Odyssey Speaker and Language Recognition
  Workshop (2020)

\bibitem{chugh2020not}
Chugh, K., Gupta, P., Dhall, A., Subramanian, R.: {Not made for each
  other-audio-visual dissonance-based deepfake detection and localization}. In:
  International Conference on Multimedia (ACM) (2020)

\bibitem{chung2018voxceleb2}
Chung, J.S., Nagrani, A., Zisserman, A.: {Voxceleb2: Deep speaker recognition}.
  In: Conference of the International Speech Communication Association
  (INTERSPEECH) (2018)

\bibitem{conti2022deepfake}
Conti, E., Salvi, D., Borrelli, C., Hosler, B., Bestagini, P., Antonacci, F.,
  Sarti, A., Stamm, M.C., Tubaro, S.: {Deepfake Speech Detection Through
  Emotion Recognition: a Semantic Approach}. In: IEEE International Conference
  on Acoustics, Speech and Signal Processing (ICASSP) (2022)

\bibitem{cozzolino2021id}
Cozzolino, D., R{\"o}ssler, A., Thies, J., Nie{\ss}ner, M., Verdoliva, L.:
  {Id-reveal: Identity-aware deepfake video detection}. In: IEEE/CVF Conference
  on Computer Vision and Pattern Recognition (CVPR) (2021)

\bibitem{desplanques2020ecapa}
Desplanques, B., Thienpondt, J., Demuynck, K.: {ECAPA-TDNN: Emphasized channel
  attention, propagation and aggregation in TDNN based speaker verification}.
  In: Conference of the International Speech Communication Association
  (INTERSPEECH) (2020)

\bibitem{deepfakesforbes}
Forbes: Deepfakes, revenge porn, and the impact on women.
  \url{https://www.forbes.com/sites/chenxiwang/2019/11/01/deepfakes-revenge-porn-and-the-impact-on-women/?sh=45b66a961f53}

\bibitem{fraudsterforbes}
Forbes: {Fraudsters Cloned Company Director’s Voice In 35\$ Million Bank
  Heist, Police Find}.
  \url{https://www.forbes.com/sites/thomasbrewster/2021/10/14/huge-bank-fraud-uses-deep-fake-voice-tech-to-steal-millions}

\bibitem{gao2021generalized}
Gao, Y., Vuong, T., Elyasi, M., Bharaj, G., Singh, R.: {Generalized Spoofing
  Detection Inspired from Audio Generation Artifacts}. In: Conference of the
  International Speech Communication Association (INTERSPEECH) (2021)

\bibitem{deepfakeguardian}
Guardian, T.: The rise of the deepfake and the threat to democracy.
  \url{https://www.theguardian.com/technology/ng-interactive/2019/jun/22/the-rise-of-the-deepfake-and-the-threat-to-democracy}

\bibitem{he2016deep}
He, K., Zhang, X., Ren, S., Sun, J.: {Deep residual learning for image
  recognition}. In: IEEE Conference on Computer Vision and Pattern Recognition
  (CVPR) (2016)

\bibitem{hosler2021deepfakes}
Hosler, B., Salvi, D., Murray, A., Antonacci, F., Bestagini, P., Tubaro, S.,
  Stamm, M.C.: {Do Deepfakes Feel Emotions? A Semantic Approach to Detecting
  Deepfakes via Emotional Inconsistencies}. In: IEEE Conference on Computer
  Vision and Pattern Recognition (CVPR) (2021)

\bibitem{hu2018squeeze}
Hu, J., Shen, L., Sun, G.: {Squeeze-and-excitation networks}. In: IEEE
  Conference on Computer Vision and Pattern Recognition (CVPR) (2018)

\bibitem{LJSpeech}
Ito, K., Johnson, L.: {The LJ Speech Dataset}.
  \url{https://keithito.com/LJ-Speech-Dataset/} (2017)

\bibitem{kamble2020advances}
Kamble, M.R., Sailor, H.B., Patil, H.A., Li, H.: Advances in anti-spoofing:
  from the perspective of {ASV}spoof challenges. APSIPA Transactions on Signal
  and Information Processing  (2020)

\bibitem{king2013}
King, S., Karaiskos, V.: {The Blizzard Challenge 2013}. In: Blizzard Challenge
  Workshop (2013)

\bibitem{li2018ictu}
Li, Y., Chang, M.C., Lyu, S.: {In ictu oculi: Exposing ai created fake videos
  by detecting eye blinking}. In: IEEE International Workshop on Information
  Forensics and Security (WIFS) (2018)

\bibitem{li2018exposing}
Li, Y., Lyu, S.: {Exposing deepfake videos by detecting face warping
  artifacts}. In: IEEE Conference on Computer Vision and Pattern Recognition
  (CVPR) (2018)

\bibitem{lieto2019hello}
Lieto, A., Moro, D., Devoti, F., Parera, C., Lipari, V., Bestagini, P., Tubaro,
  S.: {``Hello? Who Am I Talking to?" A Shallow CNN Approach for Human vs. Bot
  Speech Classification}. In: IEEE International Conference on Acoustics,
  Speech and Signal Processing (ICASSP) (2019)

\bibitem{malik2019securing}
Malik, H.: Securing voice-driven interfaces against fake (cloned) audio
  attacks. In: IEEE Conference on Multimedia Information Processing and
  Retrieval (MIPR) (2019)

\bibitem{masood2021deepfakes}
Masood, M., Nawaz, M., Malik, K.M., Javed, A., Irtaza, A.: {Deepfakes
  Generation and Detection: State-of-the-art, open challenges, countermeasures,
  and way forward}. arXiv preprint arXiv:2103.00484  (2021)

\bibitem{deepfakemimecast}
Mimecast: {Why Deepfakes are Revolutionizing the World of Phishing}.
  \url{https://www.mimecast.com/blog/deepfakes-revolutionizing-phishing}

\bibitem{nagrani2017voxceleb}
Nagrani, A., Chung, J.S., Zisserman, A.: {VoxCeleb: a large-scale speaker
  identification dataset}. In: Conference of the International Speech
  Communication Association (INTERSPEECH) (2017)

\bibitem{fakenewscientist}
{NewScientist}: {Fake faces created by AI look more trustworthy than real
  people}.
  \url{https://www.newscientist.com/article/2308312-fake-faces-created-by-ai-look-more-trustworthy-than-real-people/}

\bibitem{okabe2018attentive}
Okabe, K., Koshinaka, T., Shinoda, K.: {Attentive statistics pooling for deep
  speaker embedding}. In: Conference of the International Speech Communication
  Association (INTERSPEECH) (2018)

\bibitem{panayotov2015librispeech}
Panayotov, V., Chen, G., Povey, D., Khudanpur, S.: Librispeech: an {ASR} corpus
  based on public domain audio books. In: IEEE International Conference on
  Acoustics, Speech and Signal Processing (ICASSP) (2015)

\bibitem{pitrelli2006ibm}
Pitrelli, J.F., Bakis, R., Eide, E.M., Fernandez, R., Hamza, W., Picheny, M.A.:
  {The IBM expressive text-to-speech synthesis system for American English}.
  IEEE Transactions on Audio, Speech, and Language Processing  \textbf{14}(4),
  1099--1108 (2006)

\bibitem{ravanelli2021speechbrain}
Ravanelli, M., Parcollet, T., Plantinga, P., Rouhe, A., Cornell, S., Lugosch,
  L., Subakan, C., Dawalatabad, N., Heba, A., Zhong, J., Chou, J.C., Yeh, S.L.,
  Fu, S.W., Liao, C.F., Rastorgueva, E., Grondin, F., Aris, W., Na, H., Gao,
  Y., Mori, R.D., Bengio, Y.: {SpeechBrain: A General-Purpose Speech Toolkit}.
  arXiv:2106.04624  (2021)

\bibitem{de2021distinct}
de~Ruiter, A.: The distinct wrong of deepfakes. Philosophy \& Technology
  \textbf{34}(4),  1311--1332 (2021)

\bibitem{skerry2018towards}
Skerry-Ryan, R., Battenberg, E., Xiao, Y., Wang, Y., Stanton, D., Shor, J.,
  Weiss, R., Clark, R., Saurous, R.A.: {Towards end-to-end prosody transfer for
  expressive speech synthesis with tacotron}. In: International Conference on
  Machine Learning (ICML) (2018)

\bibitem{snyder2019speaker}
Snyder, D., Garcia-Romero, D., Sell, G., McCree, A., Povey, D., Khudanpur, S.:
  {Speaker recognition for multi-speaker conversations using x-vectors}. In:
  IEEE International Conference on Acoustics, Speech and Signal Processing
  (ICASSP) (2019)

\bibitem{snyder2018x}
Snyder, D., Garcia-Romero, D., Sell, G., Povey, D., Khudanpur, S.: {X-vectors:
  Robust DNN embeddings for speaker recognition}. In: IEEE International
  Conference on Acoustics, Speech and Signal Processing (ICASSP) (2018)

\bibitem{soxcompression}
{SoX Sound eXchange}. \url{http://sox.sourceforge.net}

\bibitem{tak2021end}
Tak, H., Patino, J., Todisco, M., Nautsch, A., Evans, N., Larcher, A.:
  {End-to-end anti-spoofing with RawNet2}. In: IEEE International Conference on
  Acoustics, Speech and Signal Processing (ICASSP) (2021)

\bibitem{deepfakenyt}
{The New York Times}: {Pennsylvania Woman Accused of Using Deepfake Technology
  to Harass Cheerleaders}.
  \url{https://www.nytimes.com/2021/03/14/us/raffaela-spone-victory-vipers-deepfake.html}

\bibitem{todisco2019asvspoof}
Todisco, M., Wang, X., Vestman, V., Sahidullah, M., Delgado, H., Nautsch, A.,
  Yamagishi, J., Evans, N., Kinnunen, T., Lee, K.A.: {ASVspoof 2019: Future
  horizons in spoofed and fake audio detection}. In: Conference of the
  International Speech Communication Association (INTERSPEECH) (2019)

\bibitem{verdoliva2020media}
Verdoliva, L.: {Media forensics and deepfakes: an overview}. IEEE Journal of
  Selected Topics in Signal Processing  \textbf{14}(5),  910--932 (2020)

\bibitem{wang2017tacotron}
Wang, Y., Skerry-Ryan, R., Stanton, D., Wu, Y., Weiss, R.J., Jaitly, N., Yang,
  Z., Xiao, Y., Chen, Z., Bengio, S., Le, Q., Agiomyrgiannakis, Y., Clark, R.,
  Saurous, R.A.: {Tacotron: Towards end-to-end speech synthesis}. In:
  Conference of the International Speech Communication Association
  (INTERSPEECH) (2017)

\bibitem{wang2018style}
Wang, Y., Stanton, D., Zhang, Y., Ryan, R.S., Battenberg, E., Shor, J., Xiao,
  Y., Jia, Y., Ren, F., Saurous, R.A.: {Style tokens: Unsupervised style
  modeling, control and transfer in end-to-end speech synthesis}. In:
  International Conference on Machine Learning (ICML) (2018)

\bibitem{wang2011channel}
Wang, Z.F., Wei, G., He, Q.H.: {Channel pattern noise based playback attack
  detection algorithm for speaker recognition}. In: IEEE International
  Conference on Machine Learning and Cybernetics (ICMLC) (2011)

\bibitem{westerlund2019emergence}
Westerlund, M.: {The emergence of deepfake technology: A review}. Technology
  Innovation Management Review  \textbf{9}(11) (2019)

\bibitem{yamagishi2021asvspoof}
Yamagishi, J., Wang, X., Todisco, M., Sahidullah, M., Patino, J., Nautsch, A.,
  Liu, X., Lee, K.A., Kinnunen, T., Evans, N., et~al.: {ASVspoof 2021:
  accelerating progress in spoofed and deepfake speech detection}. In:
  Automatic Speaker Verification and Spoofing Countermeasures Challenge (2021)

\bibitem{yang2019exposing}
Yang, X., Li, Y., Lyu, S.: {Exposing deep fakes using inconsistent head poses}.
  In: IEEE International Conference on Acoustics, Speech and Signal Processing
  (ICASSP) (2019)

\bibitem{zeinali2019but}
Zeinali, H., Wang, S., Silnova, A., Mat{\v{e}}jka, P., Plchot, O.: {BUT system
  description to voxceleb speaker recognition challenge 2019}. In: The VoxCeleb
  Challenge Workshop (2019)

\bibitem{zhang2019detecting}
Zhang, X., Karaman, S., Chang, S.F.: {Detecting and simulating artifacts in gan
  fake images}. In: IEEE International Workshop on Information Forensics and
  Security (WIFS) (2019)

\end{thebibliography}

\end{document}